\documentclass[traditabstract, longauth]{aa}
\usepackage{txfonts}
\usepackage{graphicx}
\begin{document}

\title{Probing the ATIC peak in the cosmic-ray electron spectrum with H.E.S.S.}

\author{F. Aharonian\inst{1,13}
  \and A.G.~Akhperjanian\inst{2} 
  \and G.~Anton\inst{16} 
  \and U.~Barres de Almeida\inst{8}  \thanks{supported by CAPES Foundation, Ministry of Education of Brazil}
  \and A.R.~Bazer-Bachi\inst{3}
  \and Y.~Becherini\inst{12} 
  \and B.~Behera\inst{14} 
  \and K.~Bernl\"ohr\inst{1,5} 
  \and A.~Bochow\inst{1} 
  \and C.~Boisson\inst{6}
% ...
}

\institute{
Max-Planck-Institut f\"ur Kernphysik, P.O. Box 103980, D 69029
Heidelberg, Germany
\and 
 Yerevan Physics Institute, 2 Alikhanian Brothers St., 375036 Yerevan,
Armenia
\and
Centre d'Etude Spatiale des Rayonnements, CNRS/UPS, 9 av. du Colonel Roche, BP
4346, F-31029 Toulouse Cedex 4, France
\and
Universit\"at Hamburg, Institut f\"ur Experimentalphysik, Luruper Chaussee
149, D 22761 Hamburg, Germany
\and
Institut f\"ur Physik, Humboldt-Universit\"at zu Berlin, Newtonstr. 15,
D 12489 Berlin, Germany
\and
LUTH, Observatoire de Paris, CNRS, Universit\'e Paris Diderot, 5 Place Jules Janssen, 92190 Meudon, 
France
\and
IRFU/DSM/CEA, CE Saclay, F-91191
Gif-sur-Yvette, Cedex, France
\and
University of Durham, Department of Physics, South Road, Durham DH1 3LE,
U.K.
\and
Unit for Space Physics, North-West University, Potchefstroom 2520,
    South Africa
\and
Laboratoire Leprince-Ringuet, Ecole Polytechnique, CNRS/IN2P3,
 F-91128 Palaiseau, France
\and 
Laboratoire d'Annecy-le-Vieux de Physique des Particules, 
Universit\'{e} de Savoie, CNRS/IN2P3,
9 Chemin de Bellevue - BP 110 F-74941 Annecy-le-Vieux Cedex, France
\and
Astroparticule et Cosmologie (APC), CNRS, Universite Paris 7 Denis Diderot,
10, rue Alice Domon et Leonie Duquet, F-75205 Paris Cedex 13, France
\thanks{UMR 7164 (CNRS, Universit\'e Paris VII, CEA, Observatoire de Paris)}
\and
Dublin Institute for Advanced Studies, 5 Merrion Square, Dublin 2,
Ireland
\and
Landessternwarte, Universit\"at Heidelberg, K\"onigstuhl, D 69117 Heidelberg, Germany
\and
Laboratoire de Physique Th\'eorique et Astroparticules, 
Universit\'e Montpellier 2, CNRS/IN2P3, CC 70, Place Eug\`ene Bataillon, F-34095
Montpellier Cedex 5, France
\and
Universit\"at Erlangen-N\"urnberg, Physikalisches Institut, Erwin-Rommel-Str. 1,
D 91058 Erlangen, Germany
% ...
}
\date{Received / Accepted}

\abstract{
The measurement of an excess in the cosmic-ray electron spectrum
between 300 and 800~GeV by the
ATIC experiment has - together with the PAMELA detection of a rise
in the positron fraction up to $\approx$100~GeV - 
motivated many interpretations
in terms of dark matter scenarios; alternative explanations assume a nearby electron source like a pulsar or supernova remnant.
Here we present a measurement of the cosmic-ray electron spectrum with H.E.S.S. 
starting at 340~GeV.
While the overall electron flux measured by
H.E.S.S. is consistent with the ATIC data within statistical and systematic 
errors, the H.E.S.S. data exclude a pronounced peak in the electron spectrum as 
suggested for interpretation by ATIC. The H.E.S.S. data follow a power-law 
spectrum with spectral index of $3.0\pm 0.1\mathrm{(stat.)}\pm 0.3 \mathrm{(syst.)}$, which steepens at about 1 TeV.}


\titlerunning{Probing the ATIC peak in the CR $e^\pm$ spectrum with H.E.S.S.}
\authorrunning{Aharonian et al.}
\maketitle

\section{Introduction}
Very-high-energy ($E\gtrsim100$~GeV) cosmic-ray electrons$^{1}$ 
lose their energy rapidly via 
inverse Compton scattering and synchrotron radiation 
resulting in short cooling time and hence range.
\footnotetext[1]{The term \emph{electrons} is used generically in 
the following to refer to both electrons
and positrons since most experiments do not discriminate 
between particle and antiparticle.} 
Therefore, they must come
from a few nearby sources (\cite{shen, AAV, Kobayashi}).
Recently, the ATIC collaboration reported the measurement of
an excess in the electron spectrum (\cite{atic2}). The excess appears as a peak 
in E$^3$ $\Phi$(E) where 
$\Phi$ is the differential electron flux; it can be approximated as a 
component with a power law index around 2 and a sharp cutoff around 
620~GeV. Combined with the excess in the positron
fraction measured by PAMELA (\cite{pamela}), the peak feature of the ATIC 
measurement has been interpreted in terms of a dark matter signal or 
a contribution of a nearby pulsar (e.g. \cite{Interpretation} and references
given there). 
In the case of dark matter, the structure in the electron spectrum can be 
explained as
caused by dark matter annihilation into low multiplicity
final states, while in the case of a pulsar scenario the structure 
arises from a competition between energy loss processes of pulsar 
electrons (which impose an energy cutoff depending on pulsar age) and 
energy-dependent diffusion (which favors high-energy particles in case 
of more distant pulsars).\\
The possibility to distinguish between a nearby electron source and a
dark matter
explanation with imaging atmospheric Cherenkov telescopes has been
discussed by \cite{hall}.
Imaging atmospheric Cherenkov telescopes have five orders of magnitude
larger collection areas than balloon and satellite experiments and
can therefore measure TeV electrons with excellent statistics.
Hall and Hooper assume that a structure in the electron spectrum 
should be visible even in the presence of a strong background of misidentified 
nucleonic cosmic rays. However, the assumption of a smooth background is 
oversimplified; in typical analyses the background rejection varies strongly with energy and 
without reliable control or better subtraction of the background, decisive 
results are difficult to achieve. In a recent publication, the 
High Energy Stereoscopic System (H.E.S.S.) collaboration has shown 
that such a subtraction is indeed possible, 
reporting a measurement of the electron spectrum in the range of 700~GeV to 
5~TeV~(\cite{paper1}).
\section{The low-energy extension of the H.E.S.S. electron measurement}
\begin{figure}[ht]
  \begin{center}
    \includegraphics[width=8.5cm]{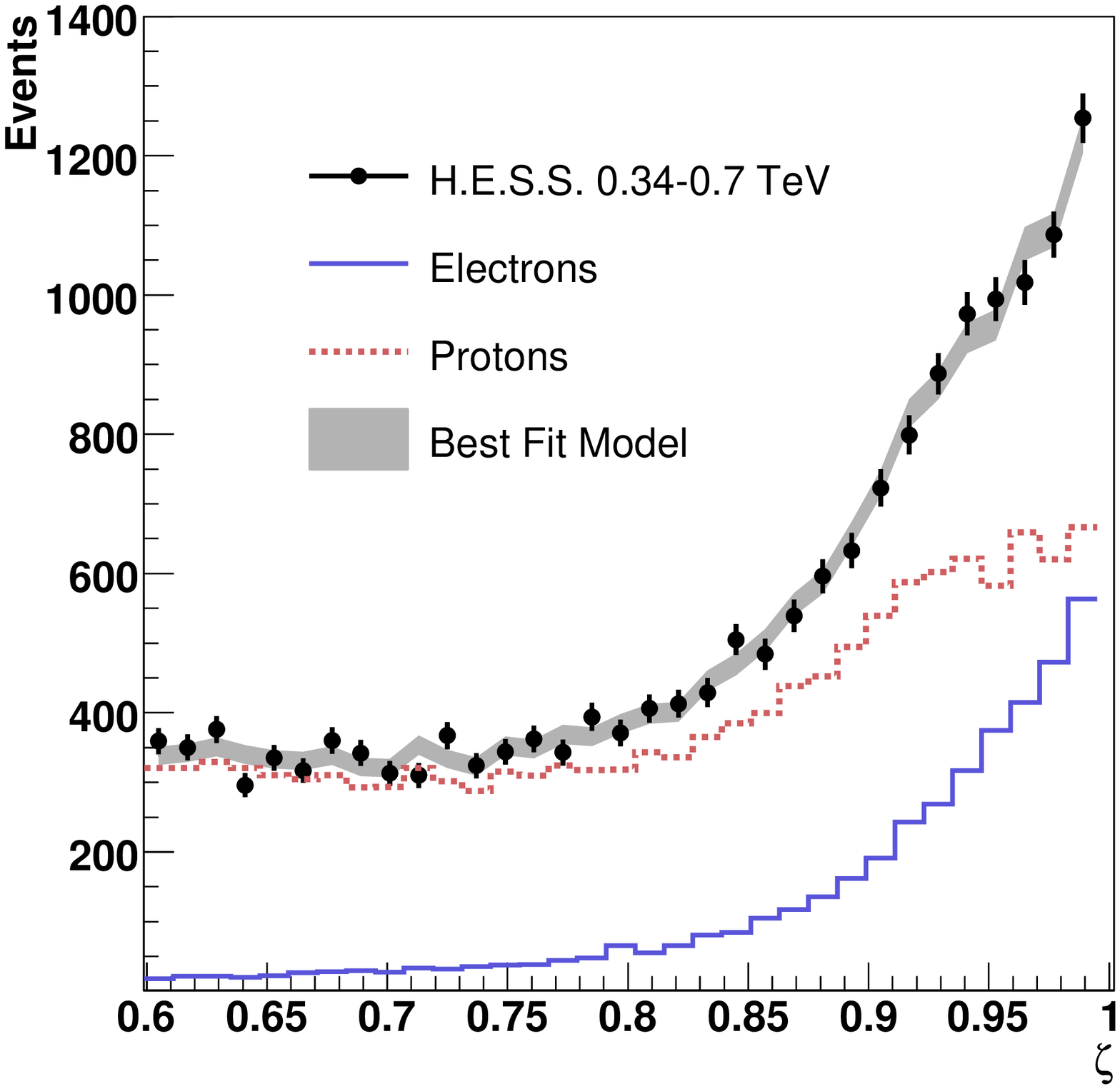}
  \vspace{-4mm}
  \end{center}
  \caption{The measured distribution of the parameter $\zeta$,
    compared with distributions for simulated protons and electrons,
    for showers with reconstructed energy between 0.34 and 0.7~TeV (the energy
    range of the extension towards lower energies compared to the analysis 
    presented in \cite{paper1}). The
    best fit model combination of electrons and protons is shown as a
    shaded band. The proton simulations use the SIBYLL hadronic
    interaction model. Distributions differ from the ones presented in Fig.~1
    of \cite{paper1} because of the energy dependence of the $\zeta$ parameter.
  }
  \label{fig1}
  \vspace{-2mm}
\end{figure}

Here an extension of the H.E.S.S. measurement towards lower energies is presented, 
partially covering the range of the reported ATIC excess.
H.E.S.S.~(\cite{HESS}) is a system
of four imaging atmospheric Cherenkov telescopes in Namibia.
While designed for the measurement of $\gamma$-ray initiated air-showers,
it can be used to measure cosmic-ray electrons as well.
The basic properties of
the analysis of cosmic-ray electrons with H.E.S.S. have been presented
in \cite{paper1}. For the analysis, data from extragalactic fields 
(with a minimum of 7$^\circ$ above or below the Galactic plane) 
are used excluding any known
or potential $\gamma$-ray source in order to avoid an almost
indistinguishable $\gamma$-ray contribution to the electron signal.
As the diffuse extragalactic $\gamma$-ray background is strongly 
suppressed by pair creation on cosmic radiation fields~(\cite{BLLacs}), 
its contribution to the measured flux can be estimated 
following \cite{BLLacs} to be less than $6\%$, 
assuming a blazar spectrum of an unbroken powerlaw
up to 3~TeV with a Gausian spectral index distribution centered at $\Gamma = -2.1$ with $\sigma_{\Gamma} = 0.35$.
For an improved rejection of the hadronic background a Random Forest
algorithm~(\cite{Forest}) is used. The algorithm uses image information
to estimate the \emph{electron likeness} $\zeta$ of each event. Since
some of the image parameters used to derive the $\zeta$ parameter are energy dependent, also $\zeta$ depends on energy.
To derive an electron spectrum, a cut   
on $\zeta$ of $\zeta > 0.6$ is applied and the number of electrons is
determined in independent energy bands by a fit of the distribution in $\zeta$
with contributions of simulated electrons and protons.
The contribution of heavier nuclei is sufficiently suppressed for $\zeta>0.6$
as not to play a role.
The result does not depend on the particular choice of $\zeta_\mathrm{min}$.
For an extension of the spectrum towards lower
energies, the analysis has been modified to improve the sensitivity at low 
energies.
In the event selection cuts, the minimum image amplitude
has been reduced from 200 to 80 photo electrons to allow 
for lower energy events.
In order to guarantee good shower reconstruction, only events with a 
reconstructed distance from the projected core position on the ground 
to the array center of less
than 100~m are included.
Additionally, only data taken 
between 2004 and 2005 are used. The reason is that the H.E.S.S. 
mirror reflectivity
degrades over time and a reduced light yield corresponds to an
increased energy threshold.
The new data and event selection reduces the event statistics but enables to lower
the analysis threshold to 340~GeV. 
The effective collection area at 340~GeV is $\approx 4 \times 10^{4}$~m$^2$.
With a live-time of 
77~hours of good quality data, a total effective exposure of 
$\approx 2.2\,\times\, 10^7$~m$^{2}$\,sr\,s is achieved at 340~GeV. 
Owing to the steepness of the electron spectrum, the measurement at lower
energies is facilitated by the comparatively higher fluxes.
The $\zeta$ distribution in the energy range of 340 to 700~GeV is shown
in Fig.~\ref{fig1}.\\
\begin{figure}[ht]
  \begin{center}
    \includegraphics[width=8.5cm]{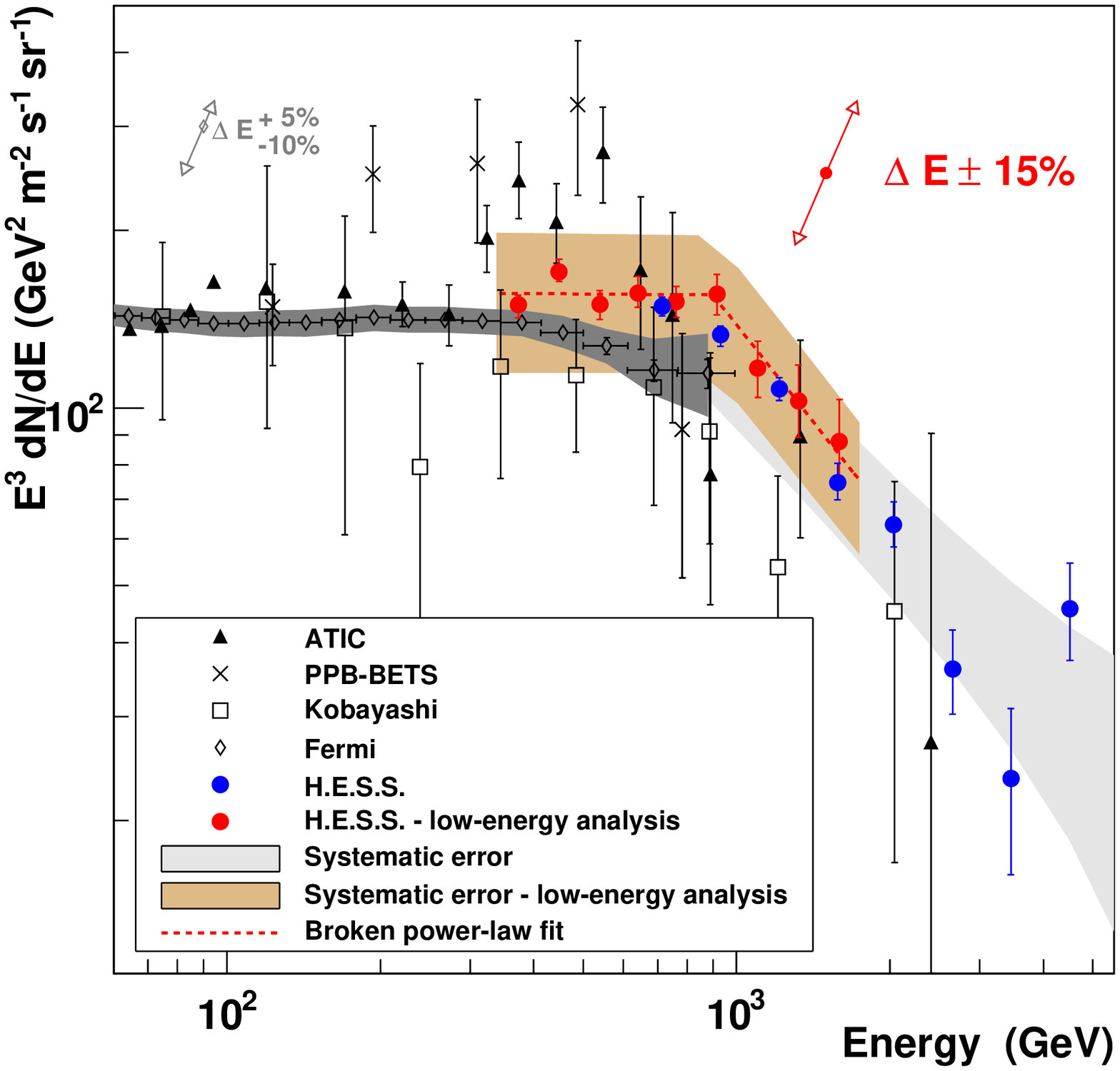}
  \vspace{-4mm}
  \end{center}
  \caption{The energy spectrum E$^3$ dN/dE of cosmic-ray electrons as measured by
    ATIC~(\cite{atic2}), PPB-BETS~(\cite{PPB_BETS}), emulsion chamber experiments~(\cite{Kobayashi}), FERMI~(\cite{fermi}) (the gray band shows the FERMI systematic uncertainty, the double arrow labeled with $+5 \% \atop -10 \%$ the uncertainty of the FERMI energy scale),
    and H.E.S.S. Previous H.E.S.S. data~(\cite{paper1}) are shown as blue points, 
    the result of the low-energy analysis presented here as red points.
    The shaded bands indicate the approximate systematic error arising from 
    uncertainties in the modeling of hadronic interactions and in the atmospheric model in the two analyses. 
    The double arrow indicates
    the effect of an energy scale shift of 15\%, the approximate 
    systematic uncertainty on the H.E.S.S. energy scale. The fit function
    is described in the text.
  }
  \label{fig2}
  \vspace{-4mm}
\end{figure}
The low-energy electron spectrum resulting from this analysis is shown in Fig.~\ref{fig2}
together with previous data of H.E.S.S. and direct measurements.
The spectrum is well described by 
a broken power law
$dN/dE =  k \cdot (E/E_{\mathrm{b}})^{-\Gamma_1}\cdot(1+(E/E_{\mathrm{b}})^{1/\alpha})^{-(\Gamma_2-\Gamma_1) \alpha}$ ($\chi^2/\mathrm{d.o.f.} = 5.6/4$, $p=0.23$) with a normalization $k=(1.5 \pm 0.1) \times 10^{-4}$~TeV$^{-1}$\,m$^{-2}$\,sr$^{-1}$\,s$^{-1}$, 
and a break energy 
$E_{\mathrm{b}} = 0.9 \pm 0.1$~TeV, where the transition between the two 
spectral indices $\Gamma_1 = 3.0\pm0.1$ and 
$\Gamma_2 = 4.1\pm0.3$ occurs. The parameter $\alpha$ denotes the sharpness of the transition, the fit prefers a sharp transition, $\alpha<0.3$.
The shaded band indicates the uncertainties in the flux normalization
that arise from uncertainties in the modeling of hadronic interactions and
in the atmospheric model. The uncertainties amount to about 30\% and are derived in the same fashion as in the 
initial paper~(\cite{paper1}), i.e. by comparison of the spectra derived from two independent data sets taken in summer and autumn 2004 for the effect of atmospheric variations and by comparison of the spectra derived using the SIBYLL and QGSJET-II hadronic interaction model for the effect of the uncertainties in the proton simulations.
The band is centered around the broken power law fit.
The systematic error on the spectral indices $\Gamma_1$, $\Gamma_2$ is $\Delta \Gamma (\mathrm{syst.})\lesssim 0.3$.
The H.E.S.S. energy scale uncertainty of $15\%$ is visualized by the
double arrow.
\section{Interpretation}
The H.E.S.S. measurement yields a smooth spectrum with a steepening towards higher energies, confirming the earlier findings above 600~GeV~(\cite{paper1}).\\
When compared to ATIC, the H.E.S.S. data show no indication of an excess and sharp cutoff in the 
electron spectrum as reported by the ATIC collaboration. Since H.E.S.S. measures the electron 
spectrum only above 340~GeV, one cannot test the rising section of the 
ATIC-reported excess.
Although different in shape, an overall consistency of the 
ATIC spectrum with the H.E.S.S. result can be obtained within the 
uncertainty of 
the H.E.S.S. energy scale of about $15\,\%$. The deviation between the ATIC 
and the H.E.S.S. data is minimal at the $20\,\%$ confidence level (assuming 
Gaussian errors for the systematic uncertainty dominating 
the H.E.S.S. measurement) when applying an upward shift of $10\,\%$ in energy to the 
H.E.S.S. data. The shift is well within the uncertainty of the H.E.S.S. 
energy scale. In this case the H.E.S.S. data overshoot the measurement 
of balloon experiments above 800~GeV, but are consistent given the 
large statistical errors from balloon experiments at 
these energies. However, the nominal H.E.S.S. data are in very good 
agreement with the high precision FERMI measurement up to 1~TeV. The combined
H.E.S.S. and FERMI measurements make a 
feature in the electron spectrum in the region of overlap of both experiments 
rather unlikely.\\
 \begin{figure}[ht]
  \begin{center}
    \includegraphics[width=8.5cm]{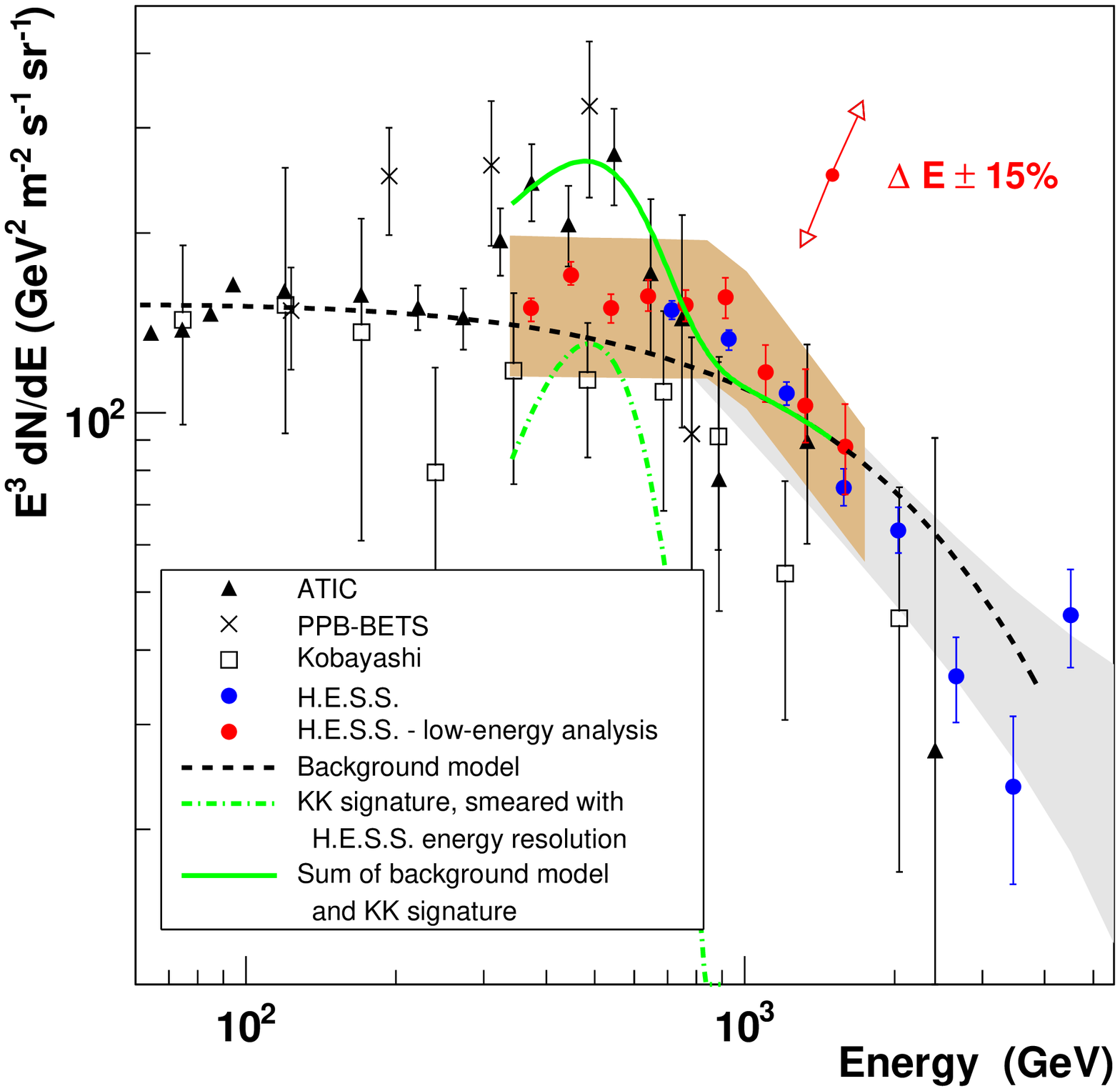}
  \vspace{-4mm}
  \end{center}
  \caption{The energy spectrum E$^3$ dN/dE of cosmic-ray electrons measured by
    H.E.S.S. and balloon experiments. Also shown are calculations for a 
    Kaluza-Klein signature in the H.E.S.S. data with a mass of 620~GeV
    and a flux as determined from the ATIC data (dashed-dotted line), the 
    background model fitted to low-energy ATIC and high-energy H.E.S.S. data (dashed line) and
    the sum of the two contributions (solid line). The shaded regions represent
    the approximate systematic error as in Fig.~\ref{fig2}.
  }
  \label{fig3}
  \vspace{-4mm}
\end{figure}
Beside comparing the H.E.S.S. measurement with ATIC and FERMI data, we also put the Kaluza-Klein (KK) interpretation suggested by \cite{atic2} to test:
A model calculation of how the therein proposed KK particle with a mass of
620~GeV and a flux approximated to fit the ATIC data 
would appear in the H.E.S.S. data is shown in Fig.~\ref{fig3}.
Here electron air showers are simulated with an energy distribution following
the energy spectrum of the KK signature presented by the ATIC collaboration. 
The simulated events and their energy are
reconstructed by the H.E.S.S. data analysis. With the use of the effective
collection area and the ``observation time'' that the number of simulations
corresponds to, the KK spectrum is obtained as it would be resolved by H.E.S.S.
Due to the limited energy resolution of about $15\%$, a sharp cutoff at the energy
of the KK mass would have been smeared out.
The residual background spectrum to a KK signal 
is modeled by a power law with exponential cutoff, which
is fitted to the low-energy ATIC data ($E < 300$~GeV) and the 
high-energy H.E.S.S. data ($E > 700$~GeV). Accordingly, our background spectrum deviates from the GALPROP 
prediction as used in \cite{atic2}. Fixing the background spectrum to most recent observational data is preferable since 
the Galactic electron spectrum at highest energies might carry the signature of nearby electron sources (\cite{pohl}) and can 
therefore differ substantially from the model calculation.
The sum of the KK signal and electron background spectrum above 340~GeV is shown as solid curve in Fig.~\ref{fig3}.
The shape of the predicted spectrum for the case of a KK signal
is not compatible with the H.E.S.S. data at the 99\% confidence level.\\
Despite superior statistics, the H.E.S.S. data do not rule out
the existence of the ATIC-reported excess owing to a possible energy scale
shift inherent to the presented measurement. Whereas compatibility with FERMI and ATIC data is confirmed, the KK scenario of \cite{atic2} cannot be easily reconciled with the H.E.S.S. measurement.
The spectrum rather exhibits a steepening
towards higher energies and is therefore compatible with conventional
electron populations of astrophysical origin within the 
uncertainties related to the injection
spectra and propagation effects.\\ 

The support of the Namibian authorities and of the University of Namibia
in facilitating the construction and operation of H.E.S.S. is gratefully
acknowledged, as is the support by the German Ministry for Education and
Research (BMBF), the Max Planck Society, the French Ministry for Research,
the CNRS-IN2P3 and the Astroparticle Interdisciplinary Programme of the
CNRS, the U.K. Science and Technology Facilities Council (STFC),
the IPNP of the Charles University, the Polish Ministry of Science and 
Higher Education, the South African Department of
Science and Technology and National Research Foundation, and by the
University of Namibia. We appreciate the excellent work of the technical
support staff in Berlin, Durham, Hamburg, Heidelberg, Palaiseau, Paris,
Saclay, and in Namibia in the construction and operation of the
equipment.


\begin{thebibliography}{99}
\bibitem[Abdo et al. 2009]{fermi} A. A. Abdo et al. 2009, Physical Review Letters, 102, 181101
\bibitem[Adriani et al. 2009]{pamela} O. Adriani et al. 2009, Nature, 458, 607
\bibitem[Aharonian et al. 1995]{AAV} F. A. Aharonian, A. M. Atoyan, H. J. V\"olk 1995, Astron. \& Astrophys., 294, L41
\bibitem[Aharonian et al. 2008]{paper1} F. A. Aharonian et al. 2008, Phys. Rev. Lett. 101, 261104
\bibitem[Breiman \& Cutler 2004]{Forest} L. Breiman \& A. Cutler, http://www.stat.berkeley.edu/ users/breiman/RandomForests/
\bibitem[Chang et al. 2008]{atic2} J. Chang et al. 2008, Nature, 456, 362 
\bibitem[Coppi \& Aharonian 1997]{BLLacs} P.S. Coppi \& F. A. Aharonian 1997, Astrophys. J., 487, L9
\bibitem[Hall \& Hooper 2008]{hall} J. Hall \& D. Hooper 2008, arXiv:0811.3362 [astro-ph] 
\bibitem[Hinton 2004]{HESS} J. A. Hinton ({\it H.E.S.S. Collaboration}) 2004, New  Astron. Rev. 48, 331 
\bibitem[Kobayashi et al. 2004]{Kobayashi} T. Kobayashi et al. 2004, Astrophys. J., 601, 340 
\bibitem[Malyshev et al. 2009]{Interpretation} D. Malyshev, I. Cholis, J. Gelfand 2009, arXiv:0903.1310v1 [astro-ph.HE]
\bibitem[Pohl \& Esposito 1998]{pohl} M. Pohl \& J. A. Esposito 1998, Astrophys. J., 507, 327
\bibitem[Shen 1970]{shen} C. S. Shen 1970, Astrophys. J. Lett., 162, L181
\bibitem[Torii et al. 2008]{PPB_BETS} S. Torii et al. 2008, arXiv:0809.0760 [astro-ph]


\end{thebibliography}
\end{document}